\documentclass[twocolumn,trackchanges]{aastex63}

\def\arcdeg{\hbox{$^\circ$}}

\def\arcsec{\hbox{$^{\prime\prime}$}}
\def\kms{\hbox{km s$^{-1}$}}
\def\VLSR{\hbox{$V_{\rm LSR}$}}
\def\Ekin{\hbox{$E_{\rm kin}$}}
\def\Pkin{\hbox{$P_{\rm kin}$}}
\def\Vexp{\hbox{$V_{\rm exp}$}}
\def\Vsys{\hbox{$V_{\rm sys}$}}
\def\texp{\hbox{$t_{\rm exp}$}}
\def\sun{\hbox{$_{\odot}$}}
\def\Leqplus{\hbox{$l\!=\!+1\fdg3$}}
\def\Leqminus{\hbox{$l\!=\!-1\fdg2$}}
\def\Mcl{\hbox{$M_{\rm cl}$}}
\def\Lcl{\hbox{$L_{\rm cl}$}}
\def\Msun{\hbox{$M_{\odot}$}}
\def\Lsun{\hbox{$L_{\odot}$}}
\newcommand{\doublecirc}{{\ooalign{$\bigcirc$\crcr\hss$\circ$\hss}}}

\received{2021 January 11}
\revised{2021 February 8}
\accepted{2021 February 12}

\submitjournal{ApJ}

\shorttitle{New Look at the Molecular Superbubble Candidate in the Galactic Center}
\shortauthors{Tsujimoto et al.}

\begin{document}

\title{New Look at the Molecular Superbubble Candidate in the Galactic Center}

%\correspondingauthor{August Muench}
%\email{greg.schwarz@aas.org, gus.muench@aas.org}

\author{Shiho Tsujimoto}
\affil{School of Fundamental Science and Technology, Graduate School of Science and Technology, Keio University, 3-14-1 Hiyoshi, Kohoku-ku, Yokohama, Kanagawa 223-8522, Japan}
\author{Tomoharu Oka}
\affiliation{School of Fundamental Science and Technology, Graduate School of Science and Technology, Keio University, 3-14-1 Hiyoshi, Kohoku-ku, Yokohama, Kanagawa 223-8522, Japan}
\affiliation{Department of Physics, Faculty of Science and Technology, Keio University, 3-14-1 Hiyoshi,  Kohoku-ku, Yokohama, Kanagawa 223-8522, Japan}
\author{Shunya Takekawa}
\affiliation{Faculty of Engineering, Kanagawa University, 3-27-1 Rokkakubashi, Kanagawa-ku, Yokohama, Kanagawa 221-8686, Japan}
\author{Yuhei Iwata}
\affiliation{School of Fundamental Science and Technology, Graduate School of Science and Technology, Keio University, 3-14-1 Hiyoshi, Kohoku-ku, Yokohama, Kanagawa 223-8522, Japan}
\author{Asaka Uruno}
\affiliation{School of Fundamental Science and Technology, Graduate School of Science and Technology, Keio University, 3-14-1 Hiyoshi, Kohoku-ku, Yokohama, Kanagawa 223-8522, Japan}
\author{Hiroki Yokozuka}
\affiliation{School of Fundamental Science and Technology, Graduate School of Science and Technology, Keio University, 3-14-1 Hiyoshi, Kohoku-ku, Yokohama, Kanagawa 223-8522, Japan}
\author{Ryosuke Nakagawara}
\affiliation{School of Fundamental Science and Technology, Graduate School of Science and Technology, Keio University, 3-14-1 Hiyoshi, Kohoku-ku, Yokohama, Kanagawa 223-8522, Japan}
\author{Yuto Watanabe}
\affiliation{School of Fundamental Science and Technology, Graduate School of Science and Technology, Keio University, 3-14-1 Hiyoshi, Kohoku-ku, Yokohama, Kanagawa 223-8522, Japan}
\author{Akira Kawakami}
\affiliation{School of Fundamental Science and Technology, Graduate School of Science and Technology, Keio University, 3-14-1 Hiyoshi, Kohoku-ku, Yokohama, Kanagawa 223-8522, Japan}
\author{Sonomi Nishiyama}
\affiliation{School of Fundamental Science and Technology, Graduate School of Science and Technology, Keio University, 3-14-1 Hiyoshi, Kohoku-ku, Yokohama, Kanagawa 223-8522, Japan}
\author{Miyuki Kaneko}
\affiliation{Department of Physics, Faculty of Science and Technology, Keio University, 3-14-1 Hiyoshi,  Kohoku-ku, Yokohama,  Kanagawa 223-8522, Japan}
\author{Shoko Kanno}
\affiliation{School of Fundamental Science and Technology, Graduate School of Science and Technology, Keio University, 3-14-1 Hiyoshi, Kohoku-ku, Yokohama, Kanagawa 223-8522, Japan}
\author{Takuma Ogawa}
\affiliation{Department of Astronomy, Graduate school of Science, The University of Tokyo, 7-3-1 Hongo Bunkyo, Tokyo 113-0033, Japan}
\affiliation{National Astronomical Observatory of Japan, 2-21-1 Osawa, Mitaka, Tokyo 181-8588, Japan}

%----------------------------------------------------------Abstract Start----------------------------------------------------------
\begin{abstract}
The \Leqplus\ region in the Galactic center is characterized by multiple shell-like structures and their extremely broad velocity widths. We revisit the molecular superbubble hypothesis for this region, based on high resolution maps of CO {\it J}=1--0, $^{13}$CO {\it J}=1--0, H$^{13}$CN {\it J}=1--0, H$^{13}$CO$^{+}$ {\it J}=1--0, SiO {\it J}=2--1, and CS {\it J}=2--1 lines obtained from the Nobeyama radio observatory 45-m telescope, as well as CO {\it J}=3--2 maps obtained from the James Clerk Maxwell telescope. We identified eleven expanding shells with total kinetic energy and typical expansion time $\Ekin\!\sim\! 10^{51.9}$ erg and $\texp\!\sim\! 10^{4.9}$ yr, respectively. In addition, the \Leqplus\ region exhibited high SiO {\it J}=2--1/H$^{13}$CN {\it J}=1--0 and SiO {\it J}=2--1/H$^{13}$CO$^{+}$ {\it J}=1--0 intensity ratios, indicating that the region has experienced dissociative shocks in the past. These new findings confirm the molecular superbubble hypothesis for the \Leqplus\ region. The nature of the embedded star cluster, which may have supplied 20--70 supernova explosions within 10$^5$ yr, is discussed. This work also show the importance of compact broad-velocity-width features in searching for localized energy sources hidden behind severe interstellar extinction and stellar contamination. 
\end{abstract}
%----------------------------------------------------------Abstract End-----------------------------------------------------------
%
%
%
%----------------------------------------------------------Introduction Start----------------------------------------------------------
\section{Introduction} \label{sec:intro}
The Galactic center contains a large quantity of warm ($T_{\rm k}\!=\!30\mbox{--}60\;{\rm K}$) and dense [$n\left({\rm H_2}\right)\!\ge\!10^4\;{\rm cm^{-3}}$] molecular gas \citep{Morris83,Paglione98}. Gases in the central molecular zone (CMZ; \citealt{Morris96}) exhibit highly turbulent and complex kinematics with large velocity dispersions \citep{Oka98}. In addition to the ubiquity of shock-origin molecules \citep{Huettemeister98, Requena-Torres06}, the highly turbulent kinematics of molecular gas in the CMZ can be attributed to the release of kinetic energy by numerous supernova (SN) explosions. It has been suggested that the instantaneous and repetitive phases of star formation have dominated continuous star formation in the CMZ in the past. Despite the abundance of dense molecular gas, star formation is currently inactive in the CMZ. The current star formation rate (SFR)$\,\sim 0.04\mbox{--}0.1\,M\sun\,{\rm yr}^{-1}$ \citep{Yusef-Zadeh09, Immer12} is at least an order of magnitude lower than that expected from the amount of dense gas in the CMZ \citep{Kruijssen14}.  

In contrast, two well-known young super star clusters in the CMZ, the Arches and Quintuplet clusters, are considered to be formed by microbursts of star formation several Myr ago \citep{Figer99a, Figer02}. Recent gamma-ray observations have revealed a pair of huge shell-like structures called ``Fermi bubbles,'' extending up to 50\arcdeg\ above and below the Galactic center. The ``Fermi bubbles'' are considered to have been formed by a pair of energetic jets from Sgr A* or a nuclear starburst in the last several tens of Myr \citep{Su10}. 
In addition, dozens of high-velocity compact clouds (HVCCs) with large kinetic energy were detected in our recent studies on the CMZ \citep{Oka98, Oka99, Oka01, Oka07, Oka12, Tanaka14, Takekawa17, Takekawa19b}. Some of the HVCCs exhibit one or more shell-like structures that may have been accelerated by a series of SN explosions that occurred in massive star clusters 10--30 Myr ago \citep{Tanaka07, Tsujimoto18}. The presence of these shells suggests that numerous micro-starbursts have occurred in the CMZ 10--30 Myr ago. The intense 6.7 keV ionized iron emission line widely detected in the CMZ via Galactic center diffuse X-ray (GCDX) was reported to be produced in hot plasma, whose possible energy source is $10^{2\mbox{--}3}$ SN explosions at a rate of $10^{-2}\mbox{--}10^{-3}\,{\rm yr^{-1}}$ \citep{Yamauchi90, Koyama07c}. This GCDX also supports the past active star formation scenario.

\begin{figure*}[htbp]
\begin{center}
\includegraphics[scale=1.0]{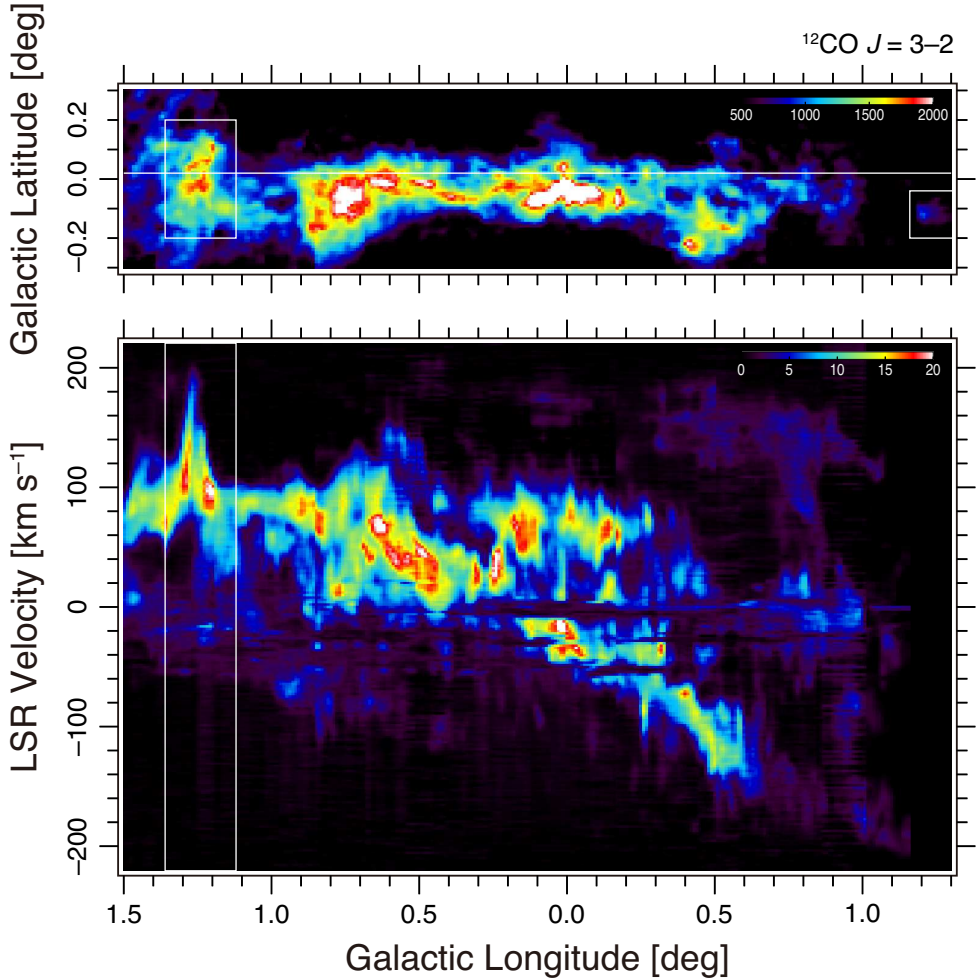}
\caption{Top: Velocity-integrated {\em l--b} map of CO {\it J}=3--2 line emission \citep{Parsons17, Eden20}. The intensity unit is K \kms. Bottom: Longitude-velocity map of CO {\it J}=3--2 emission along the white solid line in the {\em l--b} map \citep{Parsons17, Eden20}. The intensity unit is K. White squares show the areas of the \Leqplus\ and \Leqminus\ regions in the {\em l--b} and {\em l--V} maps.}
\label{fig:cmz}
\end{center}
\end{figure*}

The \Leqplus\ and \Leqminus\ regions have drawn attention as remarkably high CO {\it J}=3--2/{\it J}=1--0 intensity ratio ($R_{3\mbox{--}2/1\mbox{--}0}\ge1.5$) clumps with extremely broad velocity widths (\citealt{Oka07,Oka12}; Figure \ref{fig:cmz}). The \Leqplus\ region is associated with a large ``molecular flare'' \citep{Oka98}, which resembles the chimneys observed in starburst galaxies (e.g., \citealt{Garcia-Burillo01}).
Two expanding shells were reported in the \Leqplus\ region \citep{Oka01} before the detection of the high $R_{3\mbox{--}2/1\mbox{--}0}$ clump. These shells typically have huge kinetic energies of $\Ekin\!\sim\! 10^{52}$ erg, which corresponds to $10^{1\mbox{--}2}$ SNe. The follow-up observations of dense and shocked gas tracers in these regions have exhibited shell/arc-dominated morphology and expanding kinematics \citep{Tanaka07, Tsujimoto18}. Moreover, SiO clumps were detected at the high-velocity ends of the shells. Because SiO is a well-established probe of shocked molecular gas, we suggested that both of the \Leqplus\ and \Leqminus\ regions are proto-superbubbles containing massive ($\sim\!10^{5\mbox{--}6}\,M\sun$) star clusters. Note that the lack of a corresponding infrared observation challenges the superbubble scenario, hinting at another star formation scenario for the \Leqplus\ region (e.g., \citealt{Matsunaga20}). 

In this paper, we present newly obtained molecular line datasets of the \Leqplus\ region as parts of a large-scale survey of the CMZ (\S 2). The new CO maps show the detailed spatial and velocity distribution of this region, revealing multiple expanding shells. In addition, the distribution and kinematics of shocked gas are discussed based on the SiO line data (\S 3). Based on these datasets, we revisit the molecular superbubble scenario for the \Leqplus\ region (\S 4).  Then we summarize this work at the last section (\S 5).  The distance to the Galactic center is assumed as $D=8.3$ kpc in this paper.

%----------Figure Start----------
%----------Figure End-----------
%----------------------------------------------------------Introduction End-----------------------------------------------------------
%
%
%
%----------------------------------------------------------Data Start----------------------------------------------------------
\section{Data} \label{sec:obs}
We used the {\it J}=1--0 lines of $^{12}$CO, $^{13}$CO, H$^{13}$CN, and H$^{13}$CO$^{+}$; the {\it J}=2--1 lines of SiO and CS; and $^{12}$CO {\it J}=3--2 line data obtained by the Nobeyama radio observatory (NRO) 45-m telescope and the James Clerk Maxwell telescope (JCMT) as parts of the large-scale surveys of the Galactic CMZ or Galactic plane. We mainly analyzed the $0\fdg 24\!\times\!0\fdg4$ ($+1\fdg 12\!\le\! l\!\le\!+1\fdg 36$ and $-0\fdg 20\!\le\! b\!\le\!+0\fdg 20$) areas of each dataset. 
All reduced data were regridded onto $9\arcsec\times9\arcsec\times2\,\kms$ grids to obtain the final maps. 

\subsection{CO {\it J}=1--0 Line}
We observed the $^{12}$CO {\it J}=1--0 (115.27120 GHz) line on 19--29 January 2011, and the $^{13}$CO {\it J}=1--0 (110.20135 GHz) line on 26--31 January, 1--15 February, and 9--23 March 2016 using the NRO 45-m telescope \citep{Tokuyama19}. The target areas were set to $-0\fdg8\!\le\! l\!\le\!+1\fdg4$ and $-0\fdg35\!\le\! b\!\le\!+0\fdg35$ for the $^{12}$CO observations, and to $-1\fdg4\!\le\! l\!\le\!+1\fdg4$ and $-0\fdg35\!\le\! b\!\le\!+0\fdg35$ for the $^{13}$CO observations. The 25-beam array receiver system (BEARS; \citealt{Sunada00, Yamaguchi00}) and AC45 spectrometer \citep{Sorai00} system were employed in the $^{12}$CO line observations. Moreover, the four-beam receiver system on the 45-m telescope (FOREST; \citealt{Minamidani16}) with the spectral analysis machine on the 45-m telescope (SAM45; \citealt{Kuno11, Kamazaki12}) system were used in the $^{13}$CO line observations. The half-power beamwidths (HPBW) of the telescope with BEARS and FOREST were approximately $15\arcsec$ at 115 and 110 GHz, respectively. We used the AC45 spectrometer in the 500 MHz bandwidth (0.5 MHz resolution) mode and SAM45 spectrometer in the 1 GHz (244.14 kHz resolution) mode. The typical noise temperature ($T_{\rm sys}$) was $\sim 800$ K during the $^{12}$CO line observations, and ranged from 150 K to 300 K during the $^{13}$CO line observations. All NRO 45-m datasets were reduced on the {\it NOSTAR}\footnote{\url{http://www.nro.nao.ac.jp/~nro45mrt/html/obs/otf/export-e.html}} reduction package.
We used linear, or if necessary, the lowest degree polynomial fittings to subtract the baselines from the obtained spectra.
Less than 1\% of spectra needed non-linear baseline subtraction.
The data were spatially convolved using Bessel-Gaussian functions and resampled onto an $7\farcs5\times7\farcs5\times2\,\kms$ grid.
The temperature scale of the $^{12}$CO {\it J}=1--0 line data were determined by comparing these data with the previous $^{12}$CO {\it J}=1--0 line data \citep{Oka98}, whose intensity scale was calibrated to the radiation temperature scale of the Harvard-Smithsonian Center for Astrophysics survey \citep{Dame87}.
The $^{13}$CO {\it J}=1--0 line data in antenna temperature ($T_{\rm a}^{*}$) scale were converted to the main-beam temperature ($T_{\rm MB}$) scale by multiplying by the main-beam efficiency ($\eta_{\rm MB}$), whose value was measured to be 0.43 at 110 GHz.

\subsection{NRO 45-m Large Program}
The H$^{13}$CN {\it J}=1--0 (86.33986 GHz), H$^{13}$CO$^{+}$ {\it J}=1--0 (86.75429 GHz), SiO {\it J}=2--1 (86.84696 GHz), and CS {\it J}=2--1 (97.98096 GHz) line observations were performed in the NRO 45-m Telescope Large Program: ``Complete Imaging of the Dense and Shocked Molecular Gas in the Galactic Central Molecular Zone;'' it was approved for 2018--2019 and 2019--2020 \citep{Takekawa20}. The details of the observations and data reduction are summarized in the forthcoming paper. We mapped the area of $-1\fdg5\!\le\! l\!\le\!+1\fdg5$ and $-0\fdg25\!\le\! b\!\le\!+0\fdg25$ with the FOREST + SAM45 system in January--May 2019 and January--April 2020. The HPBW of the telescope was $\simeq 19\arcsec$ at 86 GHz. The SAM45 spectrometer was employed in the 1 GHz bandwidth (244.14 kHz resolution) mode. During the H$^{13}$CN, H$^{13}$CO, SiO, and CS line observations, $T_{\rm sys}$ ranged from 150--300 K. The obtained datasets were reduced on the {\it NOSTAR} reduction package.
We subtracted the baselines from the spectra by fitting the linear lines.
The data were spatially convolved using Bessel-Gaussian functions and resampled onto an $7\farcs5\times7\farcs5\times1\,\kms$  grid.
We converted the $T_{\rm a}^{*}$ scale into the $T_{\rm MB}$ scale with $\eta_{\rm MB}=0.49$ at 86 GHz.

\subsection{$^{12}$CO {\it J}=3--2 Line}
The JCMT Galactic plane survey team performed the $^{12}$CO {\it J}=3--2 line (345.795990 GHz) observations of the CMZ using JCMT in July--September 2013, July 2014, and March--June 2015 (14 h in total; \citealt{Parsons17, Eden20}). These data were obtained with the heterodyne array receiver program (HARP; \citealt{Buckle09}) and auto correlation spectral imaging system (ACSIS). The HPBW of the telescope was approximately $14\arcsec$ at 345 GHz. The ACSIS spectrometer was operated in the 1 GHz bandwidth (976.56 kHz) mode. During the HARP observations, $T_{\rm sys}$ typically ranged from 100--200 K. In this paper, we use the $^{12}$CO {\it J}=3--2 data for the entire \Leqplus\ region. We reduced the JCMT data with the {\it Starlink}\footnote{\url{http://starlink.eao.hawaii.edu/starlink}} software package.
The data were smoothed with Gaussian functions and resampled on a $7\farcs5\times7\farcs5\times2\,\kms$  grid to obtain the final map.
We converted the $T_{\rm a}^{*}$ scale into the $T_{\rm MB}$ scale with $\eta_{\rm MB}=0.64$ at 345 GHz.
%----------------------------------------------------------Data End-----------------------------------------------------------
%
%
%
%----------------------------------------------------------Results Start----------------------------------------------------------
\section{Results} \label{sec:result}
\subsection{Spatial Distribution} \label{subsec:spatial}
%----------Figure Start----------
\begin{figure*}[t]
\begin{center}
\includegraphics[scale=1.2]{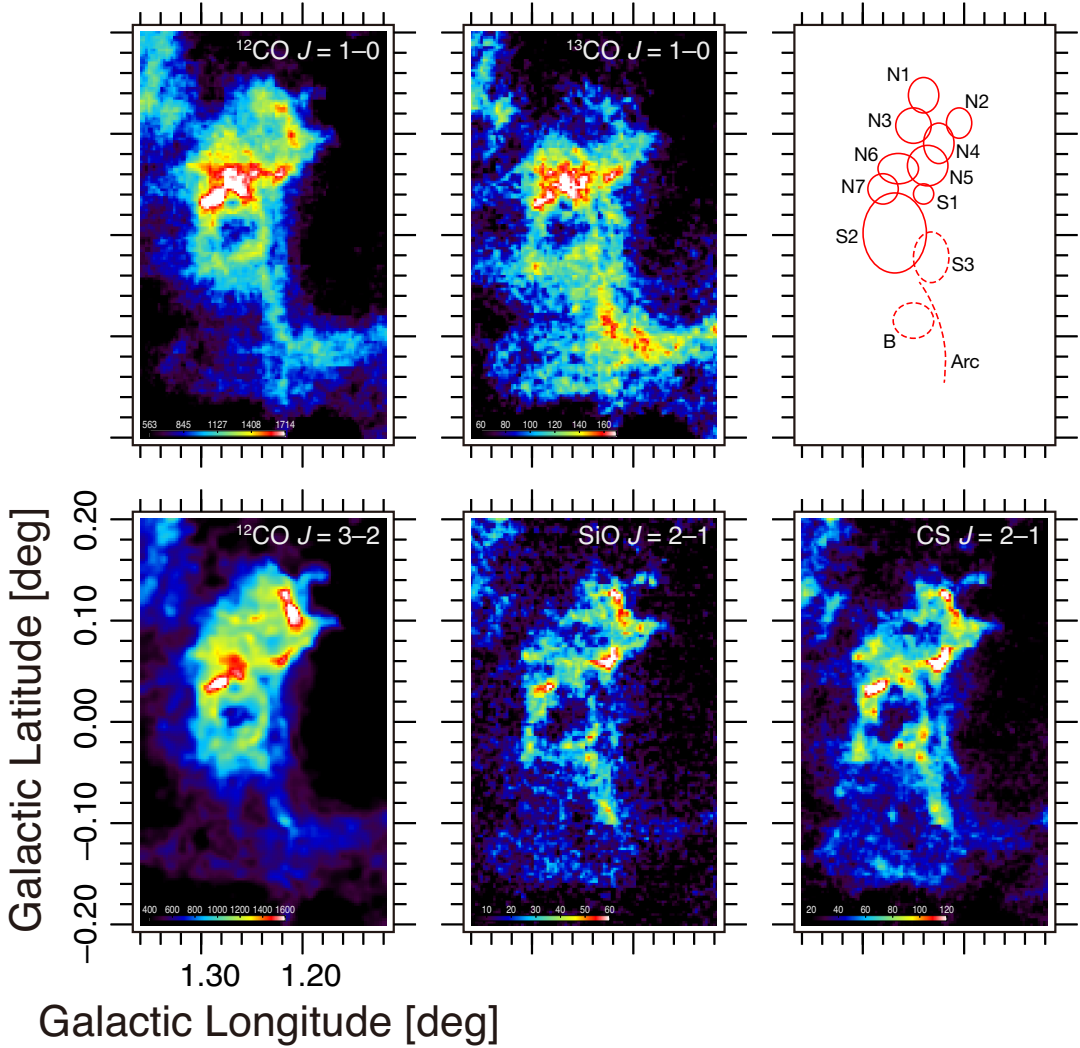}
\caption{Velocity-integrated {\em l--b} maps of $^{12}$CO and $^{13}$CO {\it J}=1--0; SiO and CS {\it J}=2--1; and $^{12}$CO {\it J}=3--2 \citep{Parsons17, Eden20} line emission. The velocity range of integration is $\VLSR=80\,\kms$ to $180\,\kms$. The intensity unit is K \kms.  The top right panel shows the locations of 11 shells identified in the $^{12}$CO {\it J}=3--2 line data. See Table \ref{tab:phys_prop} for details on the identified shells. }
\label{fig:lb-int}
\end{center}
\end{figure*}
%----------Figure End-----------
Figure \ref{fig:lb-int} shows the {\em l--b} maps of the $^{12}$CO and $^{13}$CO {\it J}=1--0; SiO and CS {\it J}=2--1; and $^{12}$CO {\it J}=3--2 line emissions integrated over a velocity range of $80\,\kms\le\VLSR\le180\,\kms$. The $^{12}$CO, SiO, and CS line emissions exhibit similar morphologies, containing two ellipses with central emission cavities centered at $(l,\,b)\sim (+1\fdg 27,\, 0\fdg 00),\,(+1\fdg 24,\, +0\fdg 10)$. Hereafter, we call these the southern and northern shells, respectively. The position of the northern shell coincides approximately with that of the ``minor shell'' described in \citet{Oka01} and the C/C$_1$ shells described in \citet{Tanaka07}. The position of the southern shell coincides with the major shell/shell A described in \citet{Oka01} and \citet{Tanaka07}. An ``arc'' structure, hereafter referred to as the Arc, was detected in the $^{12}$CO, SiO, and CS maps, which is elongated from the bottom part of the southern shell at $(l,\, b)\sim (+1\fdg 24,\, -0\fdg 04)$, reaching $(l,\,b)\sim (+1\fdg 22,\, -0\fdg 14)$. In the $^{13}$CO map, the northern shell is fainter, and the Arc is brighter and more spatially extended than those in the $^{12}$CO, SiO, and CS maps. The Arc traces the northwestern edge of shell B described in \citet{Tanaka07}.

%----------Figure Start----------
\begin{figure*}[p]
\begin{center}
\includegraphics[scale=0.90]{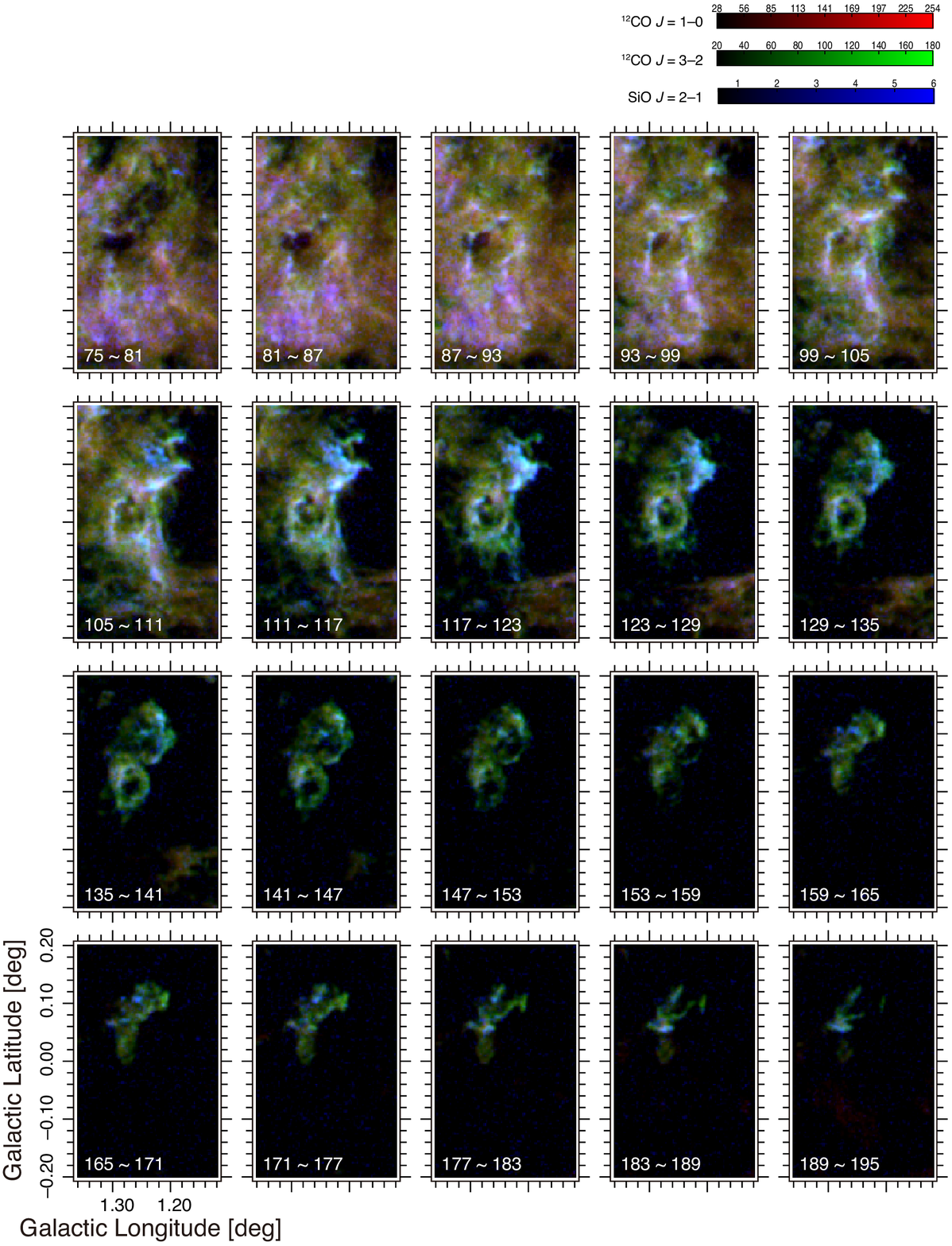}
\caption{Composite velocity channel maps of the \Leqplus\ region. Red, green, and blue indicate the velocity-integrated intensities of $^{12}$CO {\it J}=1--0, $^{12}$CO {\it J}=3--2 \citep{Parsons17, Eden20}, and SiO {\it J}=2--1 lines, respectively. The intensity unit is K \kms. The pair of numbers in the bottom left corner represents the velocity range of integration for each panel.}
\label{fig:ch}
\end{center}
\end{figure*}
%----------Figure End-----------

Figure \ref{fig:ch} shows the composite velocity channel maps of the \Leqplus\ region. 
%Red, green, and blue indicate the velocity-integrated intensities of the $^{12}$CO {\it J}=1--0, $^{12}$CO {\it J}=3--2, and SiO {\it J}=2--1 lines, respectively. 
At lower velocities ($\VLSR\le 100\,\kms$), the $^{12}$CO {\it J}=1--0 line emission dominates the color of the channel maps. At higher velocities ($\VLSR\ge100\,\kms$), where the northern and southern shells appear clearly, the $^{12}$CO {\it J}=3--2 line emission is more intense than the $^{12}$CO {\it J}=1--0 line.  The clear ellipses ($\VLSR = 117\mbox{--}153$ \kms) and deep emission holes ($\VLSR\! =\! 75\mbox{--}99$ \kms) of the southern shell are also evident. In the northern shell, the distinct arcs can also be observed at $\VLSR = 129$--159 \kms. 

Inspecting the CO {\it J}=3--2 and SiO {\it J}=2--1 datasets comprehensively, we identified 11 shell-like structures. In general, their spatial sizes change gradually with increasing velocity to form topologically closed three-dimensional structures. We named these 11 shells as follows: those around the northern shell as N1--7, those around the southern shell as S1--3, and the one adjacent to the Arc as B following \citet{Tanaka07}. The SiO {\it J}=2--1 line emission is prominent at the edges of these shells. The spatial and velocity behavior of the northern and southern shells are represented by N5 and S2, respectively. The outlines of the identified shells are shown in the top right panel of Figure \ref{fig:lb-int}. Among the 11 shells, 9 correspond to those previously identified in \citet{Oka01} and \citet{Tanaka07}. The center positions $(l,\, b)$, sizes $(\Delta l,\,\Delta b)$, and names of the corresponding shells described in previous works are listed in Table \ref{tab:phys_prop}.

%----------Figure Start----------
\begin{figure*}
\begin{center}
\includegraphics[scale=0.59]{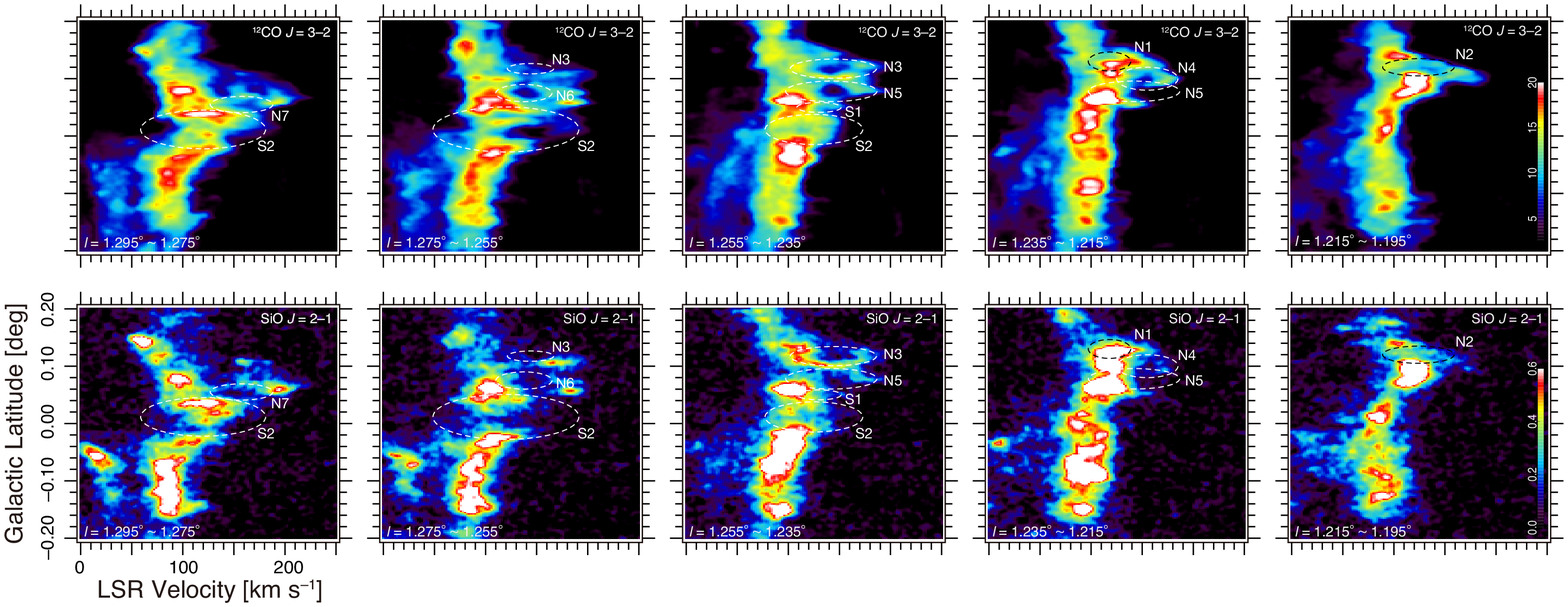}
\caption{Latitude--velocity ({\em b--V}) maps of the $^{12}$CO {\it J}=3--2 \citep{Parsons17, Eden20} and SiO {\it J}=2--1 line emissions. The averaged intensities, which were calculated for the longitude range, are shown in the bottom left corner of each panel. The intensity unit is K \arcdeg. The white-dashed ellipses denote the shells.}
\label{fig:pv}
\end{center}
\end{figure*}
%----------Figure End-----------

\subsection{Spatial-Velocity Distribution} \label{subsec:kine}
We present the {\em b--V} maps of the $^{12}$CO {\it J}=3--2 and SiO {\it J}=2--1 lines in Figure \ref{fig:pv}. The latitude--velocity ({\em b--V}) behavior of CO and SiO emissions are similar in the \Leqplus\ region, while the SiO emission favors edges and both high-velocity ends of the shells. The SiO emission is also enhanced at overlapping areas between the shells. The N1--7 and S1--2 shells appear as elliptical or arc-like structures with central cavities in these {\em b--V} maps. Such an ellipsoidal shape in the {\em l--b--V} space strongly suggests the kinematics of an ``expanding'' shell. The systemic velocity (\Vsys) and expansion velocities (\Vexp) of the shells were determined by eye, except for N3, N5, and S2 (see Section \ref{subsec:sashida}). These are listed in Table \ref{tab:phys_prop}. 

%----------Table Start----------
\begin{deluxetable*}{ccccccccccccl}
\tablecaption{Properties of Expanding Shells}\label{tab:phys_prop}
\tablehead{
\colhead{Name} & 
\colhead{\em l} & \colhead{\em b} & \colhead{\Vsys} & 
\colhead{$\Delta l$} & \colhead{$\Delta b$} & \colhead{\Vexp} & 
\colhead{${\rm log}_{10}\,M$} & \colhead{${\rm log}_{10}\,\Ekin$} & \colhead{\texp} & \colhead{${\rm log}_{10}\,\Pkin$} &
\colhead{Rank$^{*}$} & \colhead{previous denotation$^{\dag}$} \\
\colhead{} & 
\colhead{(deg)} & \colhead{(deg)} & \colhead{(\kms)} & 
\colhead{(pc)} & \colhead{(pc)} & \colhead{(\kms)} & 
\colhead{(\Msun)} & \colhead{(erg)} & \colhead{($10^4$ yr)} & \colhead{(erg s$^{-1}$)} & 
\colhead{} & \colhead{} 
 }
\startdata
N1 	&	1.2400	&	$\;\,\,\,0.1475$	&	118.0	&	4.3	&	$\;5.1$	&	$\!22$	&	3.8	&	49.5	&	$\!10.5$	&	37.0	&	$\bigcirc$		&	\,							\\	
N2	&	1.2050	&	$\;\,\,\,0.1200$	&	129.0	&	3.6	&	$\;4.3$	&	$\!35$	&	3.6	&	49.7	&	$\;5.6$	&	37.4	&	$\bigcirc$		&	\,							\\	
N3	&	1.2500	&	$\;\,\,\,0.1175$	&	144.0	&	5.1	&	$\;5.1$	&	$\;\,44.0$	&	3.9	&	50.2	&	$\;5.6$	&	37.9	&\large{$\circledcirc$}&	C$_{3}\!$				\\	
N4	&	1.2250	&	$\;\,\,\,0.1000$	&	155.0	&	4.3	&	$\;5.8$	&	$\!30$	&	3.6	&	49.6	&	$\;8.2$	&	37.1	&	$\bigcirc$		&	C$_{2}\!$				\\	
N5	&	1.2360	&	$\;\,\,\,0.0780$	&	142.0	&	5.8	&	$\;5.8$	&	$\;\,50.0$	&	4.4	&	50.8	&	$\;5.7$	&	38.5	&\large{$\circledcirc$}&	C$_{1}\!$, Minor	\\	
N6	&	1.2650	&	$\;\,\,\,0.0750$	&	137.5	&	5.8	&	$\;4.3$	&	$\!28$	&	3.7	&	49.6	&	$\;8.9$	&	37.1	&	$\bigcirc$		&	C					\\	
N7	&	1.2800	&	$\;\,\,\,0.0550$	&	157.5	&	4.3	&	$\;4.3$	&	$\!33$	&	3.8	&	49.8	&	$\;6.6$	&	37.5	&	$\bigcirc$		&	C$_{4}\!$				\\	
S1	&	1.2400	&	$\;\,\,\,0.0500$	&	130.0	&	2.9	&	$\;2.9$	&	$\!25$	&	3.7	&	49.5	&	$\;5.7$	&	37.2	&	$\bigcirc$		&	A$_{2}\!$				\\	
S2	&	1.2685	&	$\;\,\,\,0.0115$	&	119.9	&	9.1	&	$\!11.4$	&	$\;\,71.8$	&	5.1	&	51.8	&	$\;7.0$	&	39.4	&\large{$\circledcirc$}&	A, Major		\\	
S3	&	1.2325	&	$-0.0125$		&	104.0	&	5.1	&	$\;7.2$	&	$\!22$	&	4.4	&	50.1	&	$\!13.5$	&	37.5	&	$\triangle$		&	A$_{1}\!$				\\
B	&	1.2500	&	$-0.0750$		&	$\;\;94.0$	&	5.8	&	$\;5.1$	&	$\!16$	&	4.4	&	49.8	&	$\!16.6$	&	37.1	&	$\triangle$		&	B					\\\hline	
Total	&	\,		&	\,			&	\,		&	\, 	&	\,		&	\,		&	5.4	&	51.9	&	\,		&	39.5	&	\,			&	\,							\\
\enddata
%\tablecomments{References: $^{*}$ \citet{Tanaka07}, $^{\dag}$ \citet{Oka01}}
\,\\
$^{*}$\doublecirc, $\bigcirc$, and $\triangle$ are grading symbols widely used in Japan, which represent high, medium, and low evaluations, respectively.\\
$^{\dag}$``A"--``C" are the denotation in \citet{Tanaka07}.  The ``minor" and ``major" shells were first noticed by  \citet{Oka01}
\end{deluxetable*}
%----------Table End-----------
%----------------------------------------------------------Results End-----------------------------------------------------------
%
%
%
%----------------------------------------------------------Discussion Start----------------------------------------------------------
\section{Discussion} \label{sec:discuss}

\subsection{Classification of Identified Shells}
Highly complex molecular gas distribution and kinematics in the \Leqplus\ region prevents probing the shells deeply. Thus, the identifications of shells by eye are inevitably subjective. To prioritize more ``reliable'' shells in our analyses, we first ranked the 11 identified shells by their appearance into three classes: ``{\large$\circledcirc$}'' (undoubtedly expanding shells), ``{\small$\bigcirc$}'' (possibly expanding shells), and ``{$\triangle$}'' (could be expanding shells). The shells in the ``{\large$\circledcirc$}''-class exhibited clear ellipsoidal structures in the {\em l--b--V} space. The shells in the ``{\small$\bigcirc$}''-class exhibited elliptical shapes in the {\em l--b}, {\em l--V}, or {\em b--V} plane. The sets of possibly fragmented ellipses are categorized into the ``{$\triangle$}''-class. Note that the physical parameters of the lower-ranked shells generally have larger uncertainties.

\subsection{Physical Parameters} \label{subsec:phys_para}
Here, we estimate the physical parameters of the identified shells. The mass $M$ of each shell can be derived from the sum of the column density of molecular gas associated to the shell. We assume that the entire molecular gas within each {\em l--b--V} ellipsoid, which are listed in Table \ref{tab:phys_prop}, belongs to each shell. Assuming the local thermodynamic equilibrium and adopting $[^{12}{\rm CO}]/[^{13}{\rm CO}]\!=\!24$ \citep{Langer90}, $[^{12}{\rm CO}]/[{\rm H_2}]\!=\!10^{-4}$ \citep{Frerking82}, we calculated the column densities from the $^{12}$CO {\it J}=1--0 line intensity.  

The radius of each shell is defined as $R\!\equiv\! \sqrt{\Delta l\, \Delta b}/2$. Assuming that each shell expands at constant velocity \Vexp, we can calculate the expansion time by $\texp\! =\! R/\Vexp$. The kinetic energy was calculated by $\Ekin\! =\! M\Vexp^2/2$, and the kinetic power was calculated by $\Pkin=\Ekin/\texp$. The physical parameters $M$, \Ekin, \texp, and \Pkin, are also listed in Table \ref{tab:phys_prop}. 

The kinetic energy lies in the range of $\Ekin\! =\! 10^{49.5\mbox{--}51.8}$ erg. Because an SN explosion releases baryonic energy of $(1\mbox{--}3)\!\times\! 10^{50}$ erg into interstellar space \citep{Sashida13}, the kinetic energy of each {\large$\circledcirc$}-class shell corresponds to more than one SN. The expansion time is in the range of $t_{\rm exp}\!=\! (5.6\mbox{--}16.6)\times 10^4$ yr. Unlike the shells in the \Leqminus\ region \citep{Tsujimoto18}, no $t_{\rm exp}$ gradient along the spatial coordinates was observed. Except for two {$\triangle$}-class shells, the calculated $t_{\rm exp}$ coincide within a factor of 2, and \Vsys\ are confined in the range of $40$ \kms\ width. The total kinetic energy and total kinetic power of these shells amount to $10^{51.9}$ erg and $10^{39.5}$ erg s$^{-1}$, respectively.  

%----------Figure Start----------
\begin{figure*}[]
\begin{center}
\includegraphics[scale=0.7]{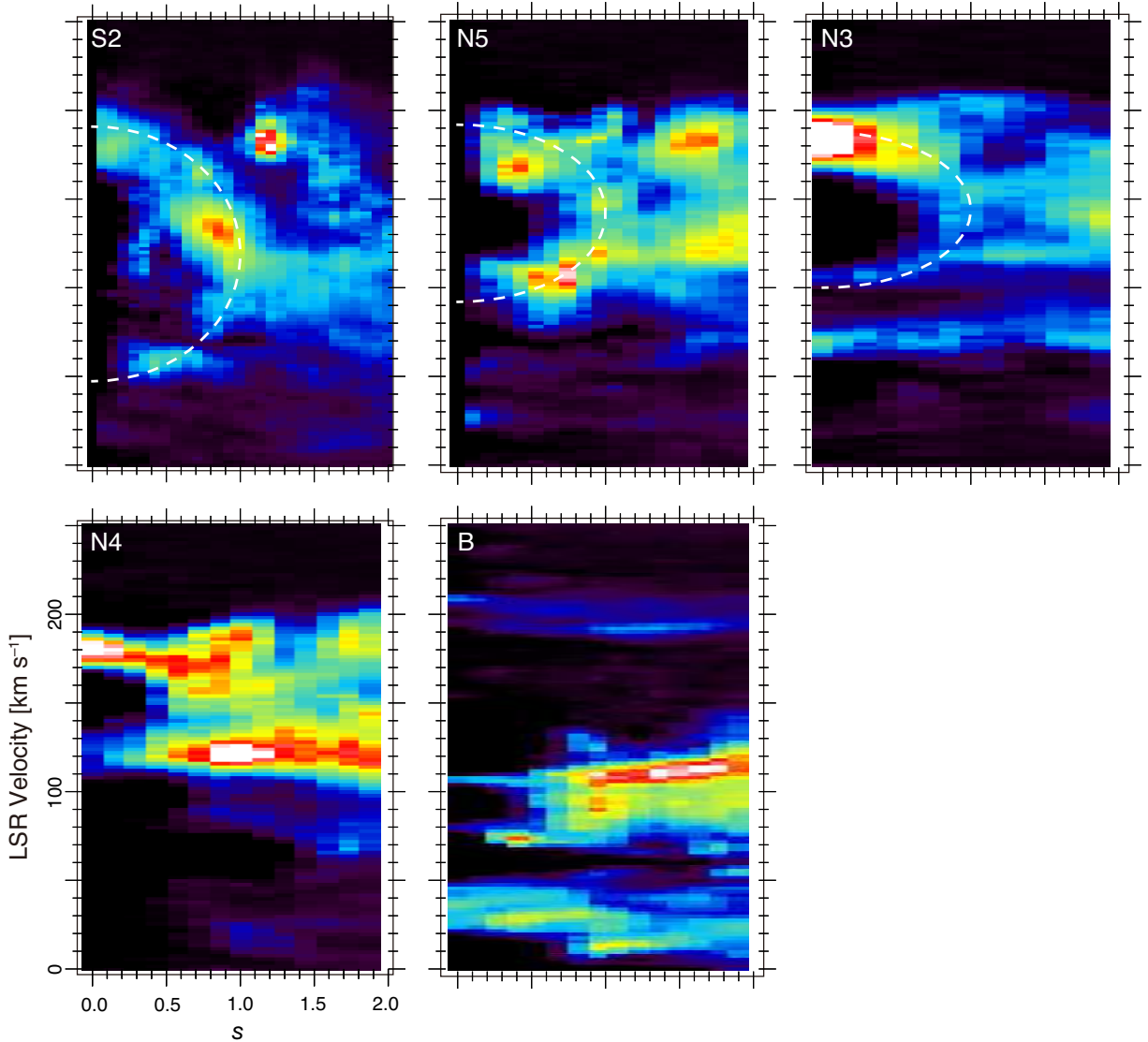}
\caption{ {\em s--V} plots of the {\large$\circledcirc$}-class shells: S2, N5, and N3; {the {\small$\bigcirc$}-class shell: N4; and the {$\triangle$}-class shell: B}. The white-dashed ellipse in each panel shows the best-fit expanding shell model in the {\em s--V} space, which provides \Vsys\ and \Vexp\ of each shell.}
\label{fig:sashida}
\end{center}
\end{figure*}
%----------Figure End-----------

\subsection{Expanding-Shell Kinematics} \label{subsec:sashida}
Here, we examine the expanding-shell kinematics for the three {\large$\circledcirc$}-class shells (S2, N5, and N3) by the {\em s--V} plot method described in \citet{Sashida13}. In this method, accurate values of \Vsys\ and \Vexp\ are determined by fitting the uniform expansion model:   
\begin{eqnarray}
\VLSR =& \Vsys+\Vexp\sqrt{1-s\left(x,\,y\right)^2} \quad & \left(s\le1\right), \label{eq:V_ellipse}\\
=& \Vsys \quad & \left(s>1\right), 
\end{eqnarray}
to the molecular line data. $s\left(x,\,y\right)$ is the normalized projected distance from the assumed center defined by 
\begin{equation}
s\left(x,\,y\right)^2 \equiv 1-\frac{1-\left(\frac{x}{a}\right)^2-\left(\frac{y}{b}\right)^2}{1+\frac{a^2-b^2}{a^2b^2}x^2}\,,
\label{eq:s}
\end{equation}
where $a$ and $b$ are the assumed semimajor and semiminor axes, respectively, of the shell in the plane of the sky. Coordinates $x$ and $y$ correspond to those along the major and minor axes of the shell, respectively. In this study, we assumed that the major axes of the shells are parallel to the Galactic longitude or latitude. We adopted the larger $\Delta l/2$ as $a$, and the smaller $\Delta b/2$ as $b$.

We used the $^{12}$CO {\it J}=3--2 data in this analysis. Before calculating the {\em s--V} plot, we applied the unsharp masking technique to the data to emphasize the ellipsoidal morphology of each shell in the {\em l--b--V} space. The spatial smoothing width was set to $0\fdg 05$, and the velocity smoothing width to 25 \kms. The center positions as well as the semimajor and semiminor axes of the shells were carefully chosen by inspecting the CO {\it J}=3--2 and SiO {\it J}=2--1 data cubes. Then, we calculated $s$ using Equation (\ref{eq:s}), to obtain the {\em s--V} plots shown in Figure \ref{fig:sashida}.  

Inner cavities can be observed in all {\em s--V} plots {for {\large$\circledcirc$}-class shells}. The ellipse in the N3 plot can be clearly observed. The cavity in the N5 plot also exhibits a clear ellipse. These observations strongly indicate the expanding-shell kinematics. By contrast, the S2 plot exhibits a distorted ellipse, and the contribution of a clump at $\VLSR\simeq 60$ \kms\ is not clear. The distortion could be due to deceleration caused by a dense clump in the northeast, or due to the overlapping of multiple expanding shells. The small clump in the S2 cavity at $\VLSR\!\simeq\! 120$ \kms\ may not be physically related to the S2 shell.  

By fitting Equation (\ref{eq:V_ellipse}) in the range of $s\le1$ to these {\em s--V} plots, we obtained \Vsys\ and \Vexp\ of S2, N5, and N3, resulting in $\Vsys\! =\! 119.9,\,142.0,\,{\rm and}\,144.0\,\kms$ and $\Vexp\! =\! 71.8,\,50.0,\,{\rm and}\,44.0\,\kms$, respectively (Table \ref{tab:phys_prop}). The N5 and N3 shells have similar $\Vsys$ and $\Vexp$, while the S2 shell has a lower $\Vsys$ and larger $\Vexp$. Despite the slight discrepancy in $\Vsys$, the {\em s--V} plot analyses clearly indicate the expanding-shell kinematics for the {\large$\circledcirc$}-class shells.  

{The {\em s--V} plots for {\small$\bigcirc$} and {$\triangle$}-classes (lower two panels in Figure \ref{fig:sashida}) also show elliptical cavities with velocity extents similar to \Vexp\ at \Vsys\, that are listed in Table \ref{tab:phys_prop}.  However, the {\em s--}extents of those cavities are less than $0.5$, while those of {\large$\circledcirc$}-class shells reach $\sim\! 1$.  These results demonstrate the difficulty in quantifying expanding-shell kinematics for lower-class shells.}

%----------Figure Start----------
\begin{figure}
\begin{center}
\includegraphics[scale=0.6]{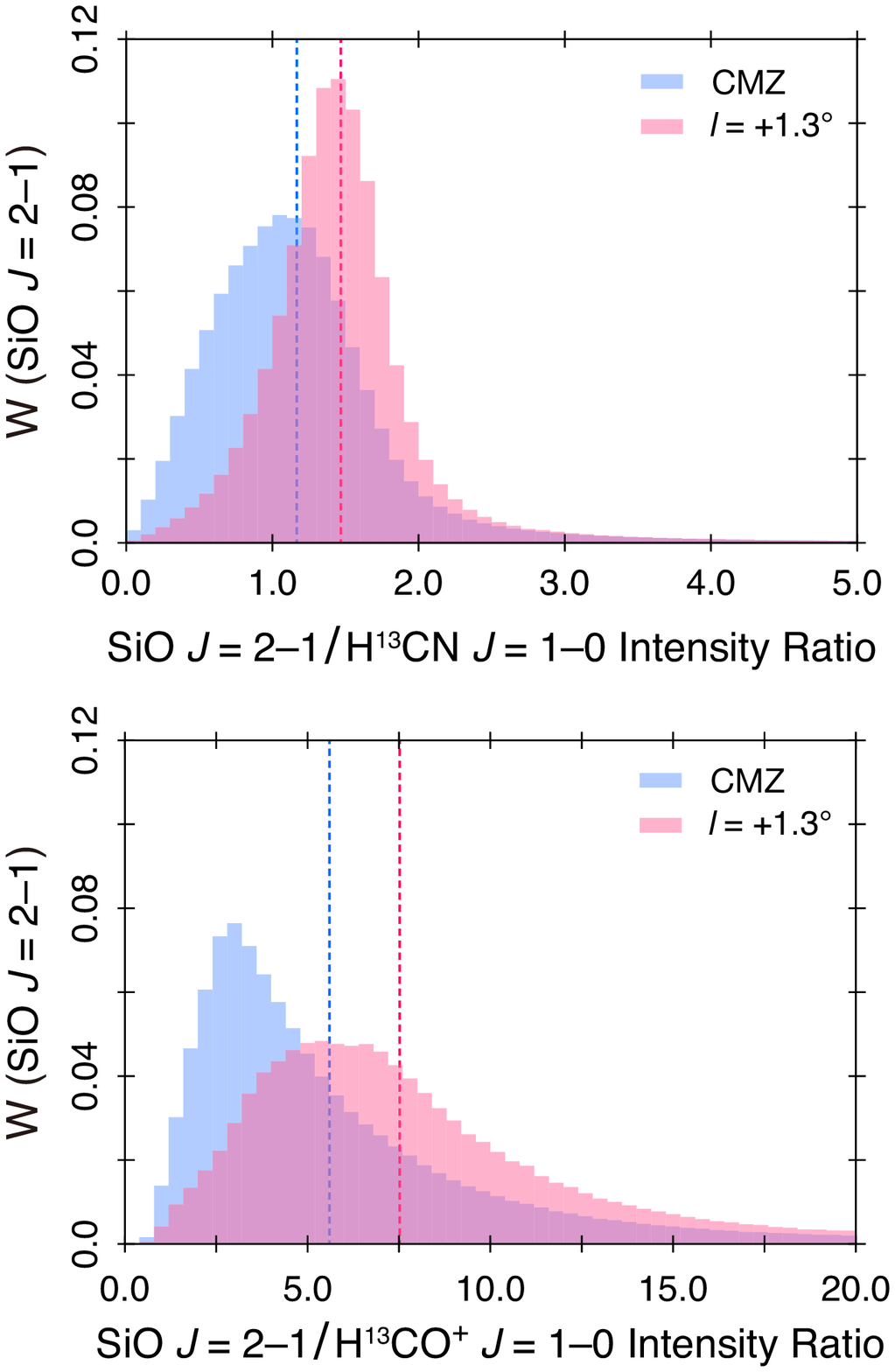}
\caption{Frequency histograms of SiO {\it J}=2--1/H$^{13}$CN {\it J}=1--0 ($R_{\rm SiO/H^{13}CN}$, top) and SiO {\it J}=2--1/H$^{13}$CO$^{+}$ {\it J}=1--0 ($R_{\rm SiO/H^{13}CO^{+}}$, bottom) intensity ratios weighted by the SiO {\it J}=2--1 intensity. Dotted lines show the mean values of the ratios of each region. Cyan and magenta represent the CMZ and \Leqplus\ region, respectively.}
\label{fig:ratio}
\end{center}
\end{figure}
%----------Figure End----------- 

\subsection{SiO Line Emission} \label{subsec:ratio}
SiO is a well-established shocked gas tracer (e.g., \citealt{Martin-Pintado92}). In the CMZ, where the molecular clouds exhibit highly turbulent kinematics, the SiO {\it J}=2--1 line is widely detected (e.g., \citealt{Tsuboi15, Takekawa20}). Nevertheless, we expect some enhancement in SiO abundance just after the passage of dissociative shock. Because the expansion times of the shells in the \Leqplus\ region is $10^{4.7\mbox{--}5.2}$ yr, its current chemical properties must have been influenced significantly by shocks.   

In Figure \ref{fig:ch}, green or cyan dominates the maps in velocities higher than $117$ \kms, where the ellipses of S2 and N5 are evident. Cyan indicates that CO {\it J}=3--2 and SiO {\it J}=2--1 emissions are present, favoring the edges of the identified shells. This indicates a higher density and/or higher SiO abundance at the edges of the shells.  

Here, we refer to the SiO {\it J}=2--1/H$^{13}$CN {\it J}=1--0 ($R_{\rm SiO/H^{13}CN}$) and SiO {\it J}=2--1/H$^{13}$CO$^{+}$ {\it J}=1--0 ($R_{\rm SiO/H^{13}CO^{+}}$) intensity ratios to examine the enhancement of SiO abundance. These ratios are considered as indicators of shock strength (e.g., \citealt{Handa06}). As the critical densities of these lines are similar ($n_{\rm cr}\!\sim\! 10^{5\mbox{--}6}$ cm$^{-3}$), they trace roughly the same spatial regions.  
Before the analyses, all datasets were smoothed with a $36\arcsec\times 36\arcsec\times 5\, \kms$ full width at half maximum Gaussian function. 

Figure \ref{fig:ratio} shows the frequency histograms of $R_{\rm SiO/H^{13}CN}$ and $R_{\rm SiO/H^{13}CO^{+}}$ weighted by the SiO {\it J}=2--1 intensity. The $R_{\rm SiO/H^{13}CN}$ distribution of the \Leqplus\ region is similar to that of the CMZ, but slightly shifted to a higher ratio. However, the $R_{\rm SiO/H^{13}CO^{+}}$ distributions are different. The \Leqplus\ region has a larger proportion of gas with higher $R_{\rm SiO/H^{13}CO^{+}}$ than the CMZ.  
The average values of $R_{\rm SiO/H^{13}CN}$ and $R_{\rm SiO/H^{13}CO^{+}}$ for the \Leqplus\ region are $1.47$ and $7.52$, both of which are larger than those for the CMZ ($1.17$ and $5.60$, respectively). 
{We performed the reduced $\chi^2$ test to examine the equality of ratio distributions between CMZ and \Leqplus\ regions.  The equality was rejected at a level of $<\! 10^{-16}$, either for $R_{\rm SiO/H^{13}CN}$ and $R_{\rm SiO/H^{13}CO^{+}}$ distributions, if we employ the typical intensity reproducibility ($8\%$) as uncertainties in the SiO intensity.}
These results suggest that the SiO abundance is enhanced in the \Leqplus\ region.  

\begin{figure*}[htbp]
\begin{center}
\includegraphics[scale=1.2]{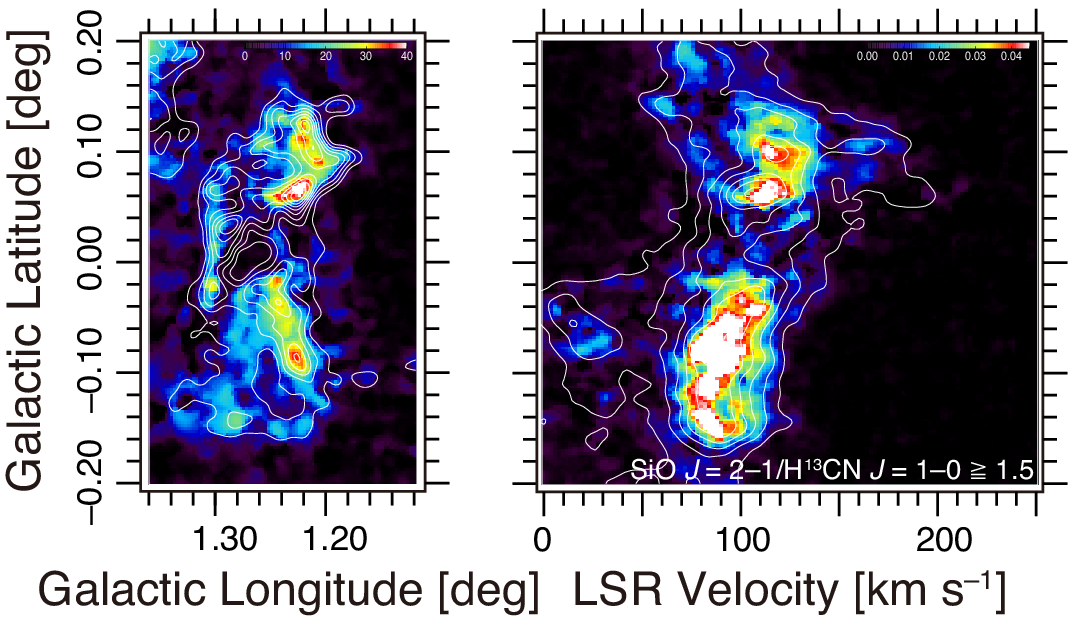}
\caption{Left: Map of SiO {\it J}=2--1 line emission integrated with $R_{\rm SiO/H^{13}CN}\ge1.5$. The intensity unit is K \kms. White contours show the total SiO {\it J}=2--1 line intensity at every 5 K \kms\ from 15 K \kms. The velocity range of the integration is $\VLSR = 80$ \kms\ to $180$ \kms. Right: Latitude--velocity map of SiO {\it J}=2--1 emission integrated with $R_{\rm SiO/H^{13}CN}\ge1.5$. The intensity unit is K \arcdeg. Contours show the $l-$integrated SiO {\it J}=2--1 emission at every 0.01 K \arcdeg. The longitude range of the integration is $l=+1\fdg 12$ to $+1\fdg  36$. }
\label{fig:highraio}
\end{center}
\end{figure*}

The spatial and velocity distributions of high $R_{\rm SiO/H^{13}CN}$ gas are shown in Figure \ref{fig:highraio}. The threshold was set to $R_{\rm SiO/H^{13}CN}=1.5$, which is roughly the average value for the \Leqplus\ region. High $R_{\rm SiO/H^{13}CN}$ gas is concentrated at the locations of shells N2--N5, B, and Arc. The most prominent concentration is found at the southwestern edge of N5. A filament of high $R_{\rm SiO/H^{13}CN}$ gas is observed along the eastern edges of S2, N7, and N6. In the latitude--velocity map, high $R_{\rm SiO/H^{13}CN}$ gas favors high intensity areas at velocities around $\VLSR=100$ \kms. In other words, the distributions of high $R_{\rm SiO/H^{13}CN}$ gas do not highlight the identified expanding shells. It roughly traces the distribution of high-density gas. This means that SiO abundance is not particularly enhanced at the identified expanding shells, while it is enhanced in the entire \Leqplus\ region.  

An explanation as to why the SiO abundance is not enhanced in the expanding shells with rather short ($\sim 10^5$ yr) expansion times may be that the abundance becomes saturated in the \Leqplus\ region. The SiO fractional abundance in the ``SiO clouds'' in the CMZ is $\sim 10^{-9}$ \citep{Martin-Pintado97}, which is similar to those in regions that experienced fast shocks. As the frequent and longstanding passages of fast shocks may have enhanced the SiO abundance in the \Leqplus\ region, an additional shock could not have increased this further.

%----------Figure Start----------
%----------Figure End-----------

\subsection{Origin of the \Leqplus\ Region} \label{sec:origin}
The broad-velocity-width nature, association of the shells/arcs and molecular flares, definite evidence for expanding kinematics in the shells, and abundance enhancement of a shock-origin molecule toward the region all support the notion that the \Leqplus\ region has been accelerated by multiple SN explosions. Such a region is called a ``molecular superbubble'' (e.g., \citealt{Oka01, Tanaka07, Tsujimoto18}). Such a superbubble must contain a young star cluster that yields SNe frequently. The absence of H$_{\rm II}$ regions indicate a cluster age of $\ge\! 10$ Myr, the main-sequence lifetime of a $15\,\Msun$ star. The presence of type-II SNe indicates that the cluster is younger than 30 Myr. This cluster age implies that the \Leqplus\ region may have experienced SNe over several Myr, which is consistent with the vertically elongated structure and widespread high $R_{\rm SiO/H^{13}CN}$ over the region.  

The total kinetic power of the \Leqplus\ region is $P_{\rm kin}\!\simeq\! 10^{39.5}$ erg s$^{-1}$. As an SN explosion releases $(1\mbox{--}3)\!\times\! 10^{50}$ erg energy into interstellar space \citep{Sashida13}, this kinetic power corresponds to an SN rate of $\sim\! 10^{-3.1}\mbox{--}10^{-3.5}$ yr$^{-1}$. Assuming that the star cluster formed instantaneously, i.e., the stellar members have the same age, the cluster mass is estimated to be $\Mcl\!\sim\! 10^{7.5}\,\Msun$, employing the Scalo initial mass function (IMF; \citealt{Scalo86}) with the low-/high-mass cutoffs and mass of the heaviest stars as $0.08\,\Msun/100\,\Msun$ and $8\,\Msun$, respectively. This unbelievably high cluster mass is a challenge that is yet to be overcome in the molecular superbubble hypothesis for the \Leqplus\ region.  

The absence of a bright infrared counterpart presents another challenge to the hypothesis. A massive cluster with an initial mass of $\Mcl\!\sim\! 10^{7.5}\,\Msun$ and age of $10\mbox{--}30$ Myr should have a total luminosity of $\Lcl\sim\! 10^{9.0}\,\Lsun$ \citep{Williams94}. However, the far infrared (100 ${\rm\mu m}$) luminosity of the \Leqplus\ region is only $L_{\rm IR}\!\sim\! 10^{6.4}\,\Lsun$, which is 2.6 orders of magnitude lower than the estimated total luminosity. This situation is similar to that encountered at the \Leqminus\ region \citep{Tsujimoto18}. This serious discrepancy between $\Lcl$ and $L_{\rm IR}$ could be partly explained by an abnormal IMF with a shallower slope and/or higher low-mass cutoff for the putative star cluster. The other types of explosions, e.g., Type Ia SNe and neutron star mergers \citep{Rosswog13} can also contribute to the kinetic power. Some extreme ideas such as the dark stellar remnant clusters should be considered in future studies.  

%----------------------------------------------------------Discussion End-----------------------------------------------------------
%
%
%
%------------------------------------------------------------Summary Start-------------------------------------------------------------
\section{Summary} \label{sec:summary}
We performed high resolution mapping observations of the H$^{13}$CN {\it J}=1--0, H$^{13}$CO$^{+}$ {\it J}=1--0, SiO {\it J}=2--1, and CS {\it J}=2--1 lines toward the \Leqplus\ region with the NRO 45-m telescope. 
By combining them with the previously obtained $^{12}$CO {\it J}=1--0, 3--2, and $^{13}$CO {\it J}=1--0 data, we attained the following results. 

\begin{enumerate}
\item We obtained high-quality maps of multiple molecular lines, including high-density and shocked gas probes in the \Leqplus\ region.

\item The \Leqplus\ region exhibits higher $R_{\rm SiO/H^{13}CN}$ and $R_{\rm SiO/H^{13}CO^{+}}$ than those in the CMZ. 
	This indicates that a strong shock has passed through this region, enhancing the SiO emission.

\item Investigating the CO {\it J}=3--2 and SiO {\it J}=2--1 maps, we identified 11 shells. 
	Of the 11 shells, 9 coincide with previously reported shells.
	\Ekin\ and \texp\ of these shells were $10^{49\mbox{--}52}$ erg and $\sim10^5$ yr, respectively.
	
\item We ranked the shells into three classes based on how apparent their elliptical shapes are in the {\it l--b--V} space, and confirmed clear expanding motion for three shells in the highest {\large$\circledcirc$}-class.
	These results support the hypothesis that the \Leqplus\ region is a superbubble.

\item The total kinetic energy of the shells are $\Ekin\sim10^{52}$ erg, and their expansion times are typically $\texp\sim10^{5}$ yr. Then, the SN rate of this region is estimated to be $\sim10^{-3.1}\mbox{--}10^{-3.5}\,{\rm yr^{-1}}$.

\item For the molecular superbubble hypothesis, there needs to be young (10--30 Myr) massive ($\Mcl\sim10^{7.5}\,\Msun$) cluster. The kinematics and morphologies of the molecular gas and energetics of this region generally support this scenario. 
\end{enumerate}

These new results reinforced the molecular superbubble hypothesis for the \Leqplus\ region, while the sharp discrepancy between the far infrared and theoretical luminosities of the embedded cluster challenges the scenario. This work also demonstrated the importance of broad-velocity-width compact molecular features in searching for localized energy sources hidden behind severe interstellar extinction and stellar contamination.  
%------------------------------------------------------------Summary End--------------------------------------------------------------
%
%
%
%----------------------------------------------------------Acknowledgements----------------------------------------------------------
\acknowledgments
The Nobeyama 45-m radio telescope is operated by Nobeyama Radio Observatory (NRO), a branch of the National Astronomical Observatory of Japan. The James Clerk Maxwell Telescope (JCMT) is operated by the East Asian Observatory on behalf of the National Astronomical Observatory of Japan, Academia Sinica Institute of Astronomy and Astrophysics, the Korea Astronomy and Space Science Institute, and the Center for Astronomical Mega-Science as well as the National Key R\&D Program of China with No. 2017YFA0402700. Additional funding support is provided by the Science and Technology Facilities Council of the United Kingdom, and participating universities in the United Kingdom and Canada. We are grateful to the staff of NRO and JCMT for their excellent support of the observations. S.Ts. acknowledges support from JSPS Grant-in-Aid for Research Fellow No. 19J10863. T.O. acknowledges support from JSPS Grant-in-Aid for Scientific Research (B) No. 15H03643. S.Ta. acknowledges support from JSPS Grant-in-Aid for Research Fellow No. 15J04405 and JSPS Grant-in-Aid for Early-Career Scientists Grant Number JP19K14768. Y.I. acknowledges support from JSPS Grant-in-Aid for Research Fellow No. JP18J20450.

\software{NOSTAR, Starlink\citep{Currie14}}

\end{document}